\documentclass[aps,prb,twocolumn,groupedaddress]{revtex4-1}

\usepackage{graphicx}
\usepackage{epsfig}
\usepackage{dcolumn}
\usepackage{bm}
\usepackage{tabularx}
\usepackage{hyperref}
\usepackage{xcolor}
\usepackage{amsmath}
\usepackage{subfigure}
\usepackage{float}
\usepackage{booktabs,dcolumn}
\usepackage{caption}
\usepackage{textcomp}
\usepackage{amssymb}
\usepackage{url}

\hypersetup{citecolor=black, filecolor= black, linkcolor= black, urlcolor= black}

\begin{document}

\title{Temperature-dependent mechanical properties of ZrC and HfC from first principles}
\author{Jin Zhang}
\affiliation{Department of Physics and Astronomy, Washington State University, Pullman, WA 99164, USA}

\author{Jeffrey M. McMahon}
\email{jeffrey.mcmahon@wsu.edu}
\affiliation{Department of Physics and Astronomy, Washington State University, Pullman, WA 99164, USA}

\begin{abstract}
In order to gain insight into the effect of elevated temperature on the mechanical performance of zirconium carbide (ZrC) and hafnium carbide (HfC), their temperature-dependent elastic constants have been systematically studied. This has been done using two different approximations, qusi-harmonic (QHA) and qusi-static (QSA) ones. The former is more accurate, but also more computational expensive. Isoentropic $C_{11}$ gradually decreases, and $C_{12}$ slightly increases for ZrC and HfC under temperature, while $C_{44}$ of both is insensitive. For both ZrC and HfC, the decline of temperature-dependent $C_{11}$ calculated from QHA is more pronounced than from QSA. Temperature effects on modulus of elasticity, Poisson's ratio, elastic anisotropy, hardness, and fracture toughness are further explored, and discussed. The results indicate that the decrease of bulk modulus \emph{B}, shear modulus \emph{G}, and Young's modulus \emph{E} approximated from QHA is more significant than from QSA at high temperatures. From room temperature to 2500 K, the theoretical decreasing slope (from QHA) of \emph{B}, \emph{G}, and \emph{E} is 0.38 GPa$\cdot$K$^{-1}$, 0.30 GPa$\cdot$K$^{-1}$, and 0.32 GPa$\cdot$K$^{-1}$, respectively, for ZrC, and 0.33 GPa$\cdot$K$^{-1}$, 0.29 GPa$\cdot$K$^{-1}$, and 0.30 GPa$\cdot$K$^{-1}$ for HfC.
\end{abstract}
\maketitle

\section{Introduction}

Hypersonic vehicles, which travel at least five times faster than the speed of sound, represent the next frontier of aircraft. Developing new materials that can withstand ultrahigh temperatures (up to 3000 degrees Celsius), thermal shock, oxidation, and corrosion at such speeds is one of the formidable challenges for design. This is especial the case for sharp nose cones and leading edges that bear the brunt of the heat. Ultrahigh-temperature ceramics (UHTCs), which exhibit a special combination of outstanding properties, such as exceptional hardness\cite{toth2014transition}, high thermal conductivity\cite{opeka1999mechanical}, good wear resistance\cite{chang2019effects}, good thermal shock resistance\cite{upadhya1997advanced}, and high melting point\cite{savino2008arc,toth2014transition,cedillos2016investigating,sheindlin2018recent}, are promising candidate materials that can meet the above performance requirements.

Among UHTCs, zirconium carbide (ZrC) and hafnium carbide (HfC) are receiving more and more attention, since they have extremely high melting points\cite{sheindlin2018recent} (3845$\pm$30 and 4255$\pm$30 K for ZrC and HfC) due to their large heat of fusion\cite{hong2015prediction} and also have good chemical inertness. The potential applications of ZrC and HfC go well beyond hypersonic flight. For example, both could be used as the thermal barrier coatings for gas turbines and ZrC is an ideal fuel-element cladding in nuclear reactors\cite{katoh2013properties}. High temperature is well-known for seriously damaging the micro- and meso-structure of a material, which results in a generalised mechanical decay. Thus, it's important to investigate the mechanical strength of ZrC and HfC at high temperatures.

The experimental study on temperature-dependent elastic constants (TDEC) of ZrC is limited to the low temperature range (4.2--298 K)\cite{chang1966low}. Elastic constants of ZrC and HfC at high temperatures are rarely reported, due to the experimental complexity of sample preparation and maintaining it at extreme temperatures. Based on density-functional theory (DFT)\cite{hohenberg1964inhomogeneous,kohn1965self}, the present study is dedicated to a systematical investigation of the elastic constants and other mechanical properties of ZrC and HfC under a wide range of temperatures.

The elastic constant discussed in this paper refers to the second-order one, which describes the linear elastic stress--strain response and wave propagation in solids\cite{born1954dynamical,wallace1998thermodynamics}. Several theoretical approaches\cite{varshni1970temperature,shrivastava1980temperature,garber1975theory,singh2005temperature,wang2010first,shang2010temperature,shao2012temperature} have been proposed to study the TDEC of single crystals. These can mainly be classified into two categories: the empirical formula\cite{varshni1970temperature,shrivastava1980temperature,garber1975theory,singh2005temperature,liu2007elastic,li2019temperature} and the method combining first-principles calculations with finite-strain continuum elasticity theory\cite{thurston1964physical,wallace1970thermoelastic,zhao2007first}. Since the applicability of the empirical models or formulas is limited to either some specific materials or to the cubic crystals, the latter method, which can be used for crystals with arbitrary symmetry, is the most commonly utilized technique in calculating TDEC. In this method, the elastic constants at a temperature are determined as the second-order strain derivatives of the Helmholtz free energy change induced by a homogeneous strain. Generally, this energy under temperature can be obtained from three different methods: empirical Debye models\cite{anderson1995equations,moruzzi1988calculated}, lattice dynamics\cite{togo2015first}, and first-principles molecular dynamics\cite{vovcadlo2007ab}. Here, the widely used lattice dynamics method is considered.

Two methods of estimating the TDEC can be originally derived from the lattice dynamics\cite{wang2010first,shang2010temperature}. One is called the quasi-harmonic approximation (QHA), and the other is the quasi-static approximation (QSA). These two approaches will be described in the methods section. Compared with the QHA, the computational cost of QSA is much lower, but it is also more approximate. A recent theoretical report on ZrC\cite{jing2018phase} is calculated using QSA method due to its cheap cost. QSA method is also widely used for calculating the TDEC of other materials\cite{papadimitriou2015ab,papadimitriou2015abtwo,liu2015first,chong2016effects,shang2012effects}. Therefore, another motivation of this study is to make a comparison between the QHA and QSA methods in estimating TDEC and other temperature-dependent mechanical properties.

\section{Methods}

\subsection{Temperature-dependent elastic constants}

Based on finite-strain continuum elasticity theory\cite{wallace1972}, the internal energy at a volume of a crystal under a general strain can be expressed by expanding that for \emph{E}(\emph{V}) of the deformed crystal with respect to the strain tensor (defined below), in terms of a Taylor series, as
\begin{align}\label{eq1}
E(V) &= E(V_{0})+V_{0}\sum_{m,n}\sigma_{mn}\delta_{mn} \nonumber \\
     &+\frac{V_{0}}{2!}\sum_{m,n,k,l}C_{mnkl}^{S}\delta_{mn}\delta_{kl}+\cdots
\end{align}
where \emph{E}(\emph{V}$_{0}$) is the ground state energy, \emph{V} is the volume of the strained crystal, \emph{V}$_{0}$ is the volume of the unstrained crystal, $\sigma_{mn}$ are the elements of stress tensor, and $\delta_{mn}$ are those of strain tensor. Note that each strain corresponds to a different deformation tensor. Then, the second-order isentropic elastic constants $C_{mnkl}^{S}$ can be expressed as
\begin{equation}\label{eq2}
C_{mnkl}^{S}=\frac{\partial^2E}{\partial \delta_{mn} \partial\delta_{kl}}~~~.
\end{equation}
This is the general equation to calculate the ground-state elastic constants from DFT. The Helmholtz free energy $\emph{F}$ can also be expanded in terms of a Taylor series\cite{brugger1964thermodynamic,davies1974effective,zhao2007first}
\begin{align}\label{eq3}
F(V,T) & =F(V_{0},T)+V_{0}\sum_{m,n}\sigma_{mn}\delta_{mn} \nonumber \\
       & +\frac{V_{0}}{2!}\sum_{m,n,k,l}C_{mnkl}^{T}\delta_{mn}\delta_{kl}+\cdots ~~~
\end{align}
where \emph{T} is the temperature. Thus, the second-order isothermal elastic constants $C_{mnkl}^{T}$ at constant temperature is derived as
\begin{equation}\label{eq4}
C_{mnkl}^{T}=\frac{\partial^2F}{\partial \delta_{mn} \partial\delta_{kl}}~~~.
\end{equation}
Note the Voigt notation\cite{brugger1964thermodynamic} is used for the tensor indices to write $C_{mnkl}$ as $C_{ij}$ in the subsequent discussion. In Eq.\ (\ref{eq4}), \emph{F}(T) is usually approximated by
\begin{equation}\label{eq5}
F(T)= E_{s}+F_{el}(T)+F_{vib}(T)
\end{equation}
where \emph{E}$_{s}$ is the energy of a static lattice at 0 K, \emph{F}$_{el}$ is the thermal electronic free-energy arising from electronic excitations, which can be determined integration over the electronic density of state through the Fermi-Dirac distribution\cite{wasserman1996thermal} (\emph{F}$_{el}$ is neglected in the following calculations, since the non-zero electronic density at the Fermi level of ZrC and HfC), and \emph{F}$_{vib}$ is the lattice vibrational energy contribution, which can be obtained by the partition function of lattice vibration.

Therefore, to calculate the elastic constants under temperature, one needs to perform three steps\cite{wang2010first}:
1. Apply homogeneous deformation on the optimized structure to obtain serval strained ones. Then, a sets of distortions is applied to each strained structure with a small strain. The elastic stiffness constants under temperature are extracted by fitting the strain-Helmholtz free energy relationship [Eq.\ (\ref{eq3})]. 2. Using the first-principles quasiharmonic approach, predict the temperature at which these strained structures in the first step correspond to. 3. Based on the results from above two steps, the relationship between temperature and elastic constants is obtained using interpolation.
This process of calculating TDEC is named as the quasi-harmonic approximation (QHA). For simplicity of calculation, the contribution of vibrational free energy \emph{F}$_{vib}$ \emph{could} be neglected in calculating the Helmholtz free energy [Eq.\ \ref{eq5}] during step 1; this approximation is named as the quasi-static approximation (QSA).

Most elastic stiffness coefficients are measured by the resonance ultrasound spectroscopy, where elastic waves induced deformation in the crystal can be viewed as an isoentropic process. Therefore, to make direct comparison to the experimental elastic constants, $C_{ij}^{T}$ need to be converted to $C_{ij}^{S}$ (i.e., which are often measured). This can be achieved through the relationship\cite{davies1974effective},

\begin{equation}\label{eq7}
C_{ij}^{S}(T)=C_{ij}^{T}(T)+\frac{TV\lambda_{i} \lambda_{j} }{C_{v}}
\end{equation}
where $C_{v}$ is the isochoric heat capacity and the coefficients $\lambda$ are calculated as
\begin{equation}\label{eq8}
\lambda_{i}=-\sum_{j}\alpha_{j}C_{ij}^{T}(T)
\end{equation}
where $\alpha_{j}$ is the thermal expansion tensor.

\subsection{Details of first-principles and phonon calculations}

These above three steps for calculating TDEC are implemented in {\verb"thermo"\verb"_pw"}\cite{thermopwmanual} package combined with the \emph{ab initio} DFT calculations\cite{hohenberg1964inhomogeneous,kohn1965self}. The latter was performed within the projector-augmented wave (PAW) pseudopotentials\cite{blochl1994projector,kresse1999ultrasoft}, as implemented in the {\sc Quantum ESPRESSO} package\cite{giannozzi2009quantum,QE-2017}. For the exchange and correlation terms in the electron--electron interaction, the generalized-gradient approximation (GGA) of Perdew--Burke--Eruzerhof (PBE)\cite{perdew1996generalized} was used. The kinetic-energy cutoff for the plane wave basis set was chosen as 120 and 140 Ry for ZrC and HfC, respectively. $\mathbf{k}$-point grids were based on 10$\times$10$\times$10 Monkhorst-Pack (MP) meshes. For Brillouin zone integration, the first-order Methfessel--Paxton method\cite{methfessel1989high} was used with a smearing width of 0.02 Ry. These choices give a convergence in energy to less than 1 meV/atom. Density functional perturbation theory (DFPT)\cite{baroni2001phonons} was used for phonon calculations, as (also) implemented within {\sc Quantum ESPRESSO}. A 4$\times$4$\times$4 $\mathbf{q}$-point mesh was used to produce the dynamical matrices for the phonon calculations.

\section{Results and discussions}

\begin{center}
\begin{figure}[htp]
\includegraphics[angle=0,width=0.55\linewidth]{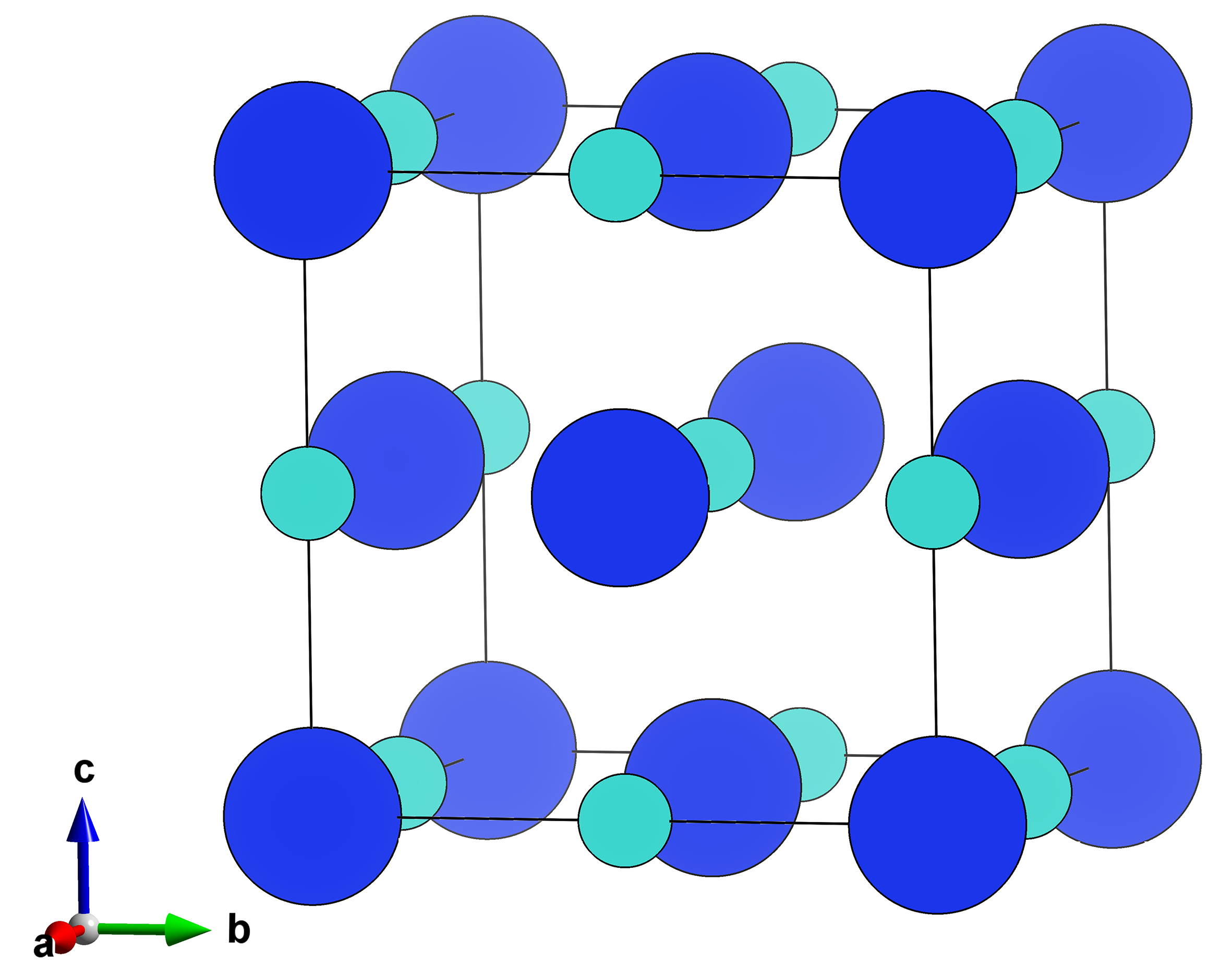}
\caption{\label{Fig1} (Color online) The crystal structure of NaCl-type ZrC and HfC. Large balls are Zr or Hf atoms, and samll balls are C atoms.}
\end{figure}
\end{center}

Both polycrystalline ZrC and HfC have an NaCl-type structure, as shown in Fig.\ \ref{Fig1}. The optimized lattice-parameter of ZrC is calculated as 4.724 {\AA}{} and that of HfC is 4.646 {\AA}{} at 298 K. These are in good agreement with the experimental data (4.693 {\AA}{} for ZrC\cite{lawson2007thermal} and 4.644 {\AA}{} for HfC\cite{smith1970phonon}).

\begin{center}
\begin{figure*}
\includegraphics[angle=0,width=0.85\linewidth]{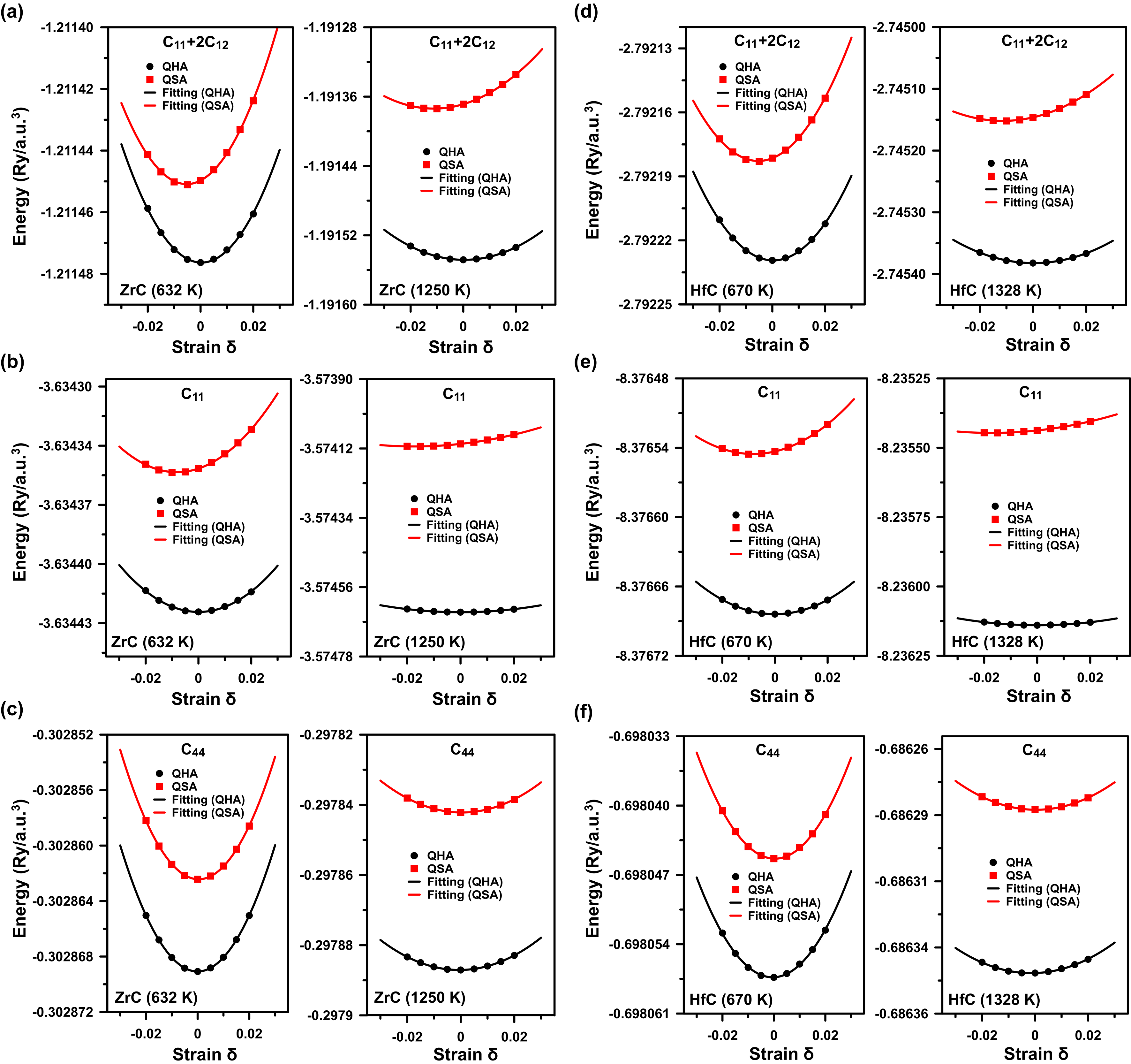}
\caption{\label{Fig2} (Color online) Calculated relationships between Helmholtz free energy density and strains for (a)--(c) ZrC and (d)--(f) HfC. The dot and square symbols represent the result calculated from the quasi-harmonic (QHA) and quasi-static (QSA) approximations, respectively. The lines are fitted by polynomial.}
\end{figure*}
\end{center}

There are three elastic constants $C_{11}$, $C_{12}$, and $C_{44}$ that need to be calculated for the cubic crystal structure. The number of strained structures obtained from homogeneous deformation in step 1 (in the methods section) were chosen as 17. The sets of distortions were selected as ($\delta$, $\delta$, $\delta$, 0, 0, 0), (0, 0, $\delta$, 0, 0, 0), and (0, 0, 0, $\delta$, $\delta$, $\delta$) on each (homogeneously) strained-structure to, respectively, calculate $C_{11}$+$C_{12}$, $C_{11}$, and $C_{44}$ with a small strain $\delta$ varying from -0.02 to 0.02 in steps of 0.005. Fig.\ \ref{Fig2} shows the calculated Helmholtz energy densities against these strain curves for two selected (homogeneously) strained-structures, which correspond to two different temperatures. The exact values of these temperatures can be obtained from the first-principles quasiharmonic approach (as mentioned in step 2, in the methods section).

\begin{center}
\begin{figure*}
\includegraphics[angle=0,width=0.75\linewidth]{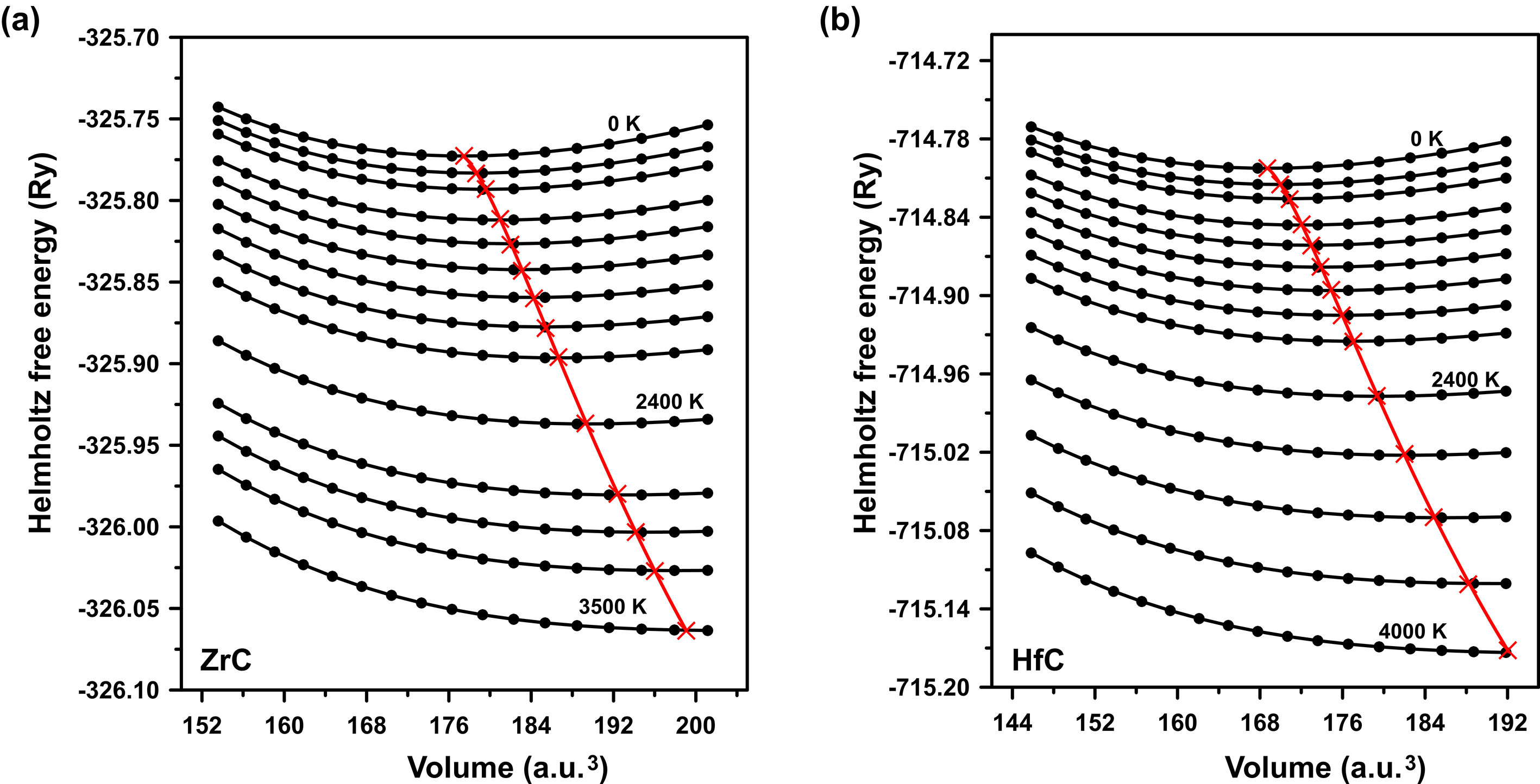}
\caption{\label{Fig3} (Color online) Volume-dependent Helmholtz free energies of (a) ZrC and (b) HfC. Red lines connect the equilibrium volume with the minimum free energy at every temperature; Thus, the red crosses indicate the equilibrium volume at each temperature.}
\end{figure*}
\end{center}

By fourth-order polynomial fitting the Helmholtz free energy densities to strain [see Fig. \ref{Fig2}(a)-(f)], the second-order coefficient can be obtained, and it equals to the corresponding linear combination of elastic constants. The elastic constants at a given temperature can be calculated by solving simultaneously these three linear equations. For both ZrC [Fig. \ref{Fig2}(a)--(c)] and HfC [Fig. \ref{Fig2}(d)--(f)], the parabola at high temperatures is ``wider" than that at low one, indicating the elastic constants become smaller. The difference between the line with filled dots for QHA and squares for QSA shows that the latter method can overestimate the value of Helmholtz free energy, which becomes greater at higher temperature.

\begin{center}
\begin{figure*}
\includegraphics[angle=0,width=0.75\linewidth]{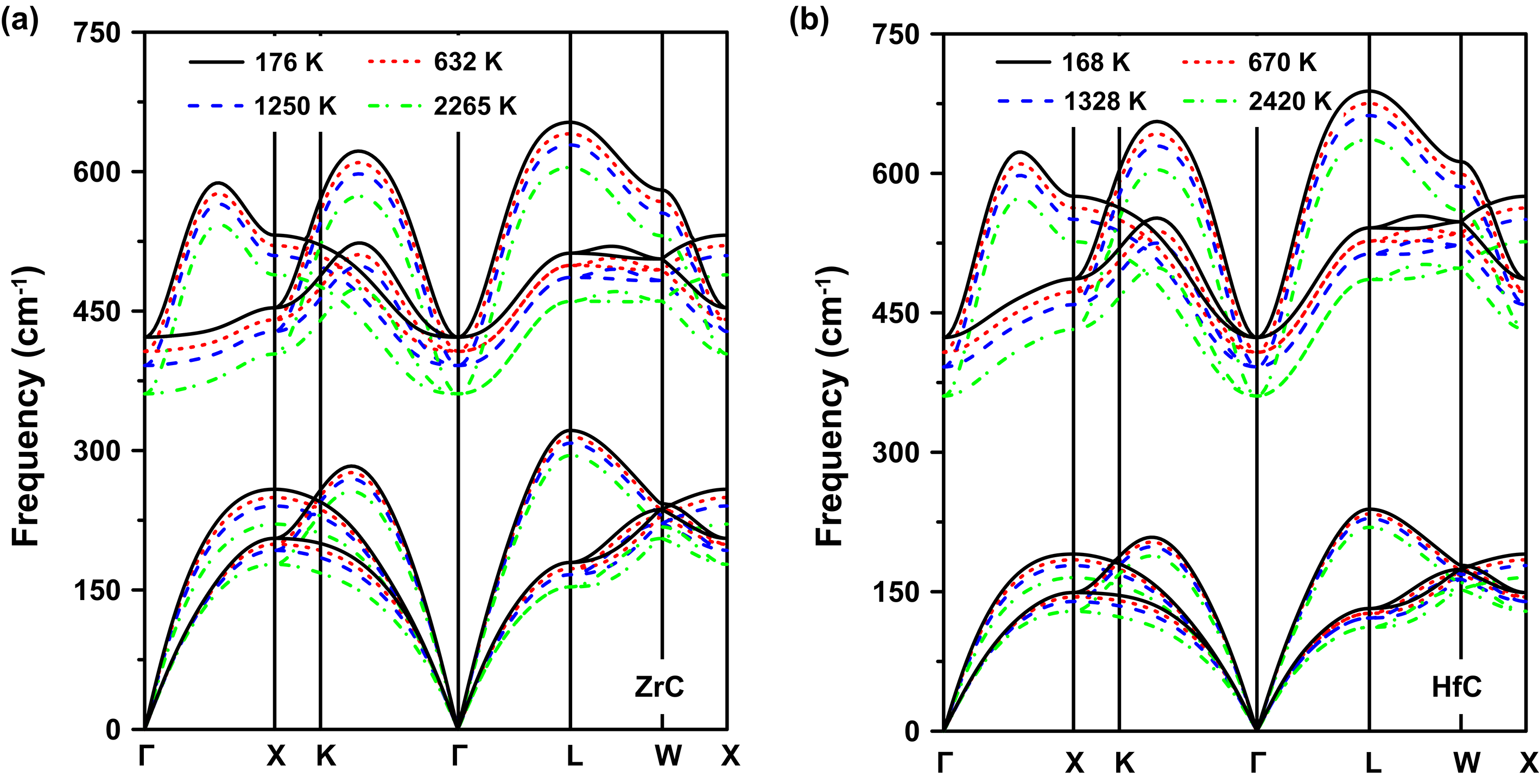}
\caption{\label{Fig4} (Color online) Phonon dispersion relations of (a) ZrC and (b) HfC along high-symmetry directions over the Brillouin zone at several temperatures.}
\end{figure*}
\end{center}

\begin{center}
\begin{figure*}
\includegraphics[angle=0,width=0.75\linewidth]{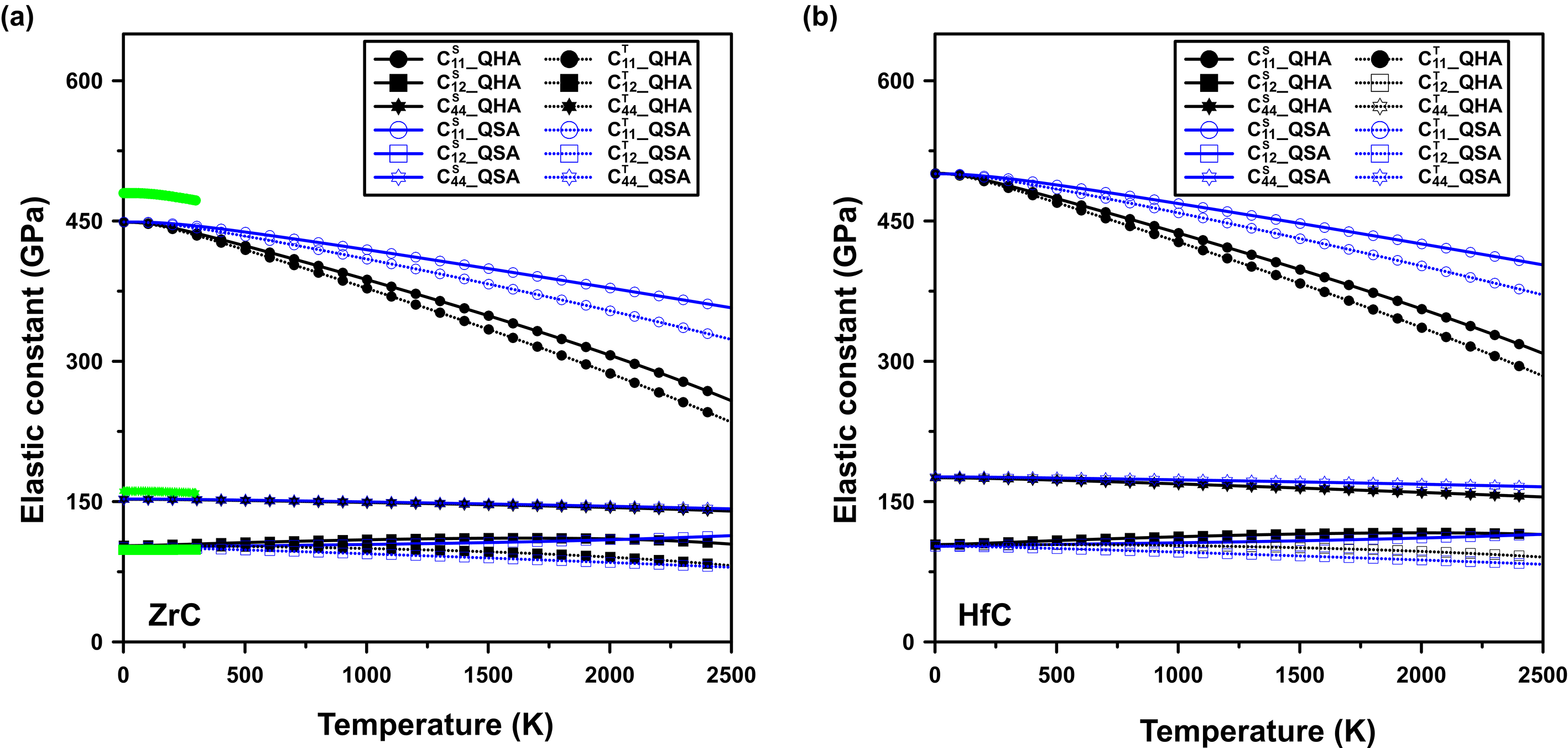}
\caption{\label{Fig5} (Color online) Temperature-dependent isoentropic (solid lines) and isothermal (dashed lines) elastic constants for (a) ZrC and (b) HfC. Also shown in (a) is experimental data\cite{chang1966low} (green). The lines with filled and open symbols represent calculations by the QHA and QSA approaches, respectively.}
\end{figure*}
\end{center}

\begin{center}
\begin{figure*}
\includegraphics[angle=0,width=0.75\linewidth]{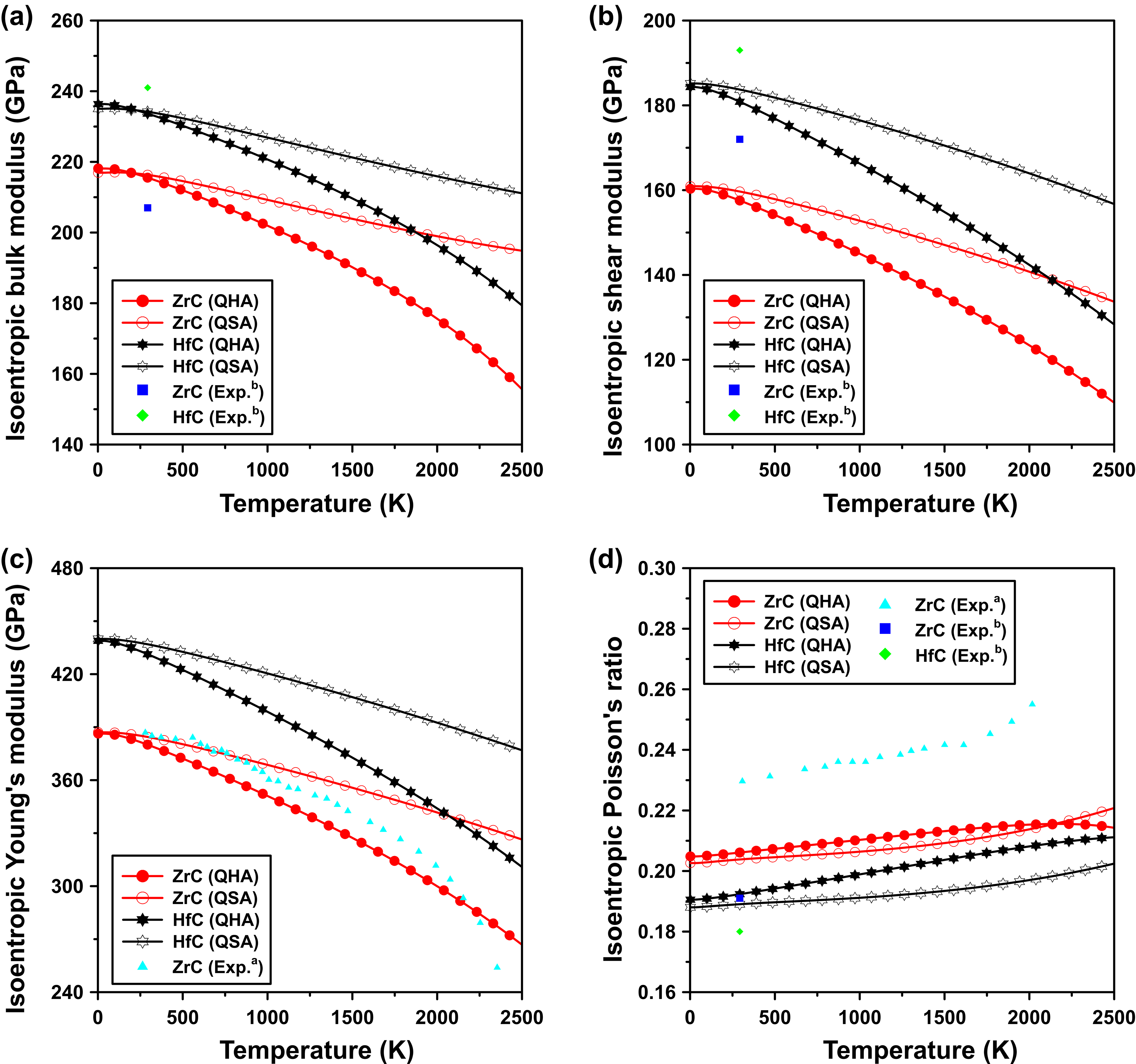}
\caption{\label{Fig6} (Color online) Temperature-dependent (a) bulk modulus \emph{B}, (b) shear modulus \emph{G}, (c) Young's modulus \emph{E}, and (d) Poisson's ratio $\upsilon$ of ZrC and HfC, based on the isoentropic $C_{ij}^{S}$ and Hill's approach\cite{hill1952elastic}. The lines with filled and open symbols represent elastic moduli based on the isoentropic $C_{ij}^{S}$ calculated by the QHA and QSA approaches, respectively. Exp.$^{a}$ is from Ref. \onlinecite{baranov1973temperature}; Exp.$^{b}$ (at 293 K) is from Ref. \onlinecite{pierson1996carbides}.}
\end{figure*}
\end{center}

Fig.\ \ref{Fig3} shows the Helmholtz free energy at finite temperature for different crystal volumes, where the 17 volume points chosen were the same as those in step 1. These calculations were performed using the first-principles quasiharmonic approach (step 2, in the methods section). The equilibrium volume at each temperature was determined by fitting the data to the Vinet equation of state\cite{vinet1989universal}.

Phonon dispersion curves of ZrC and HfC at different temperatures are shown in Fig. \ref{Fig4}. There are no unstable branches with imaginary vibrational frequencies, indicating that the structures are dynamically stable at these temperatures. The Born elastic stability criteria\cite{born1940stability,mouhat2014necessary} for the cubic crystal system($C_{11}$-$C_{12}$$>$0, $C_{11}$+2$C_{12}$$>$0, $C_{44}$$>$0) was applied for checking the mechanical stability of ZrC and HfC. It can be seen that the temperature effect on phonon frequencies is discernible, and the overall trend is that they decrease as temperature increases. This is understandable, since the phonon frequency typically decreases with increasing volume.

The QHA and QSA methods were used to estimate the isoentropic elastic constants $C_{ij}^{S}$ of ZrC and HfC as a function of temperature. $C_{ij}^{S}$ and the corresponding isothermal elastic constants $C_{ij}^{T}$ drawn for comparison are shown in Fig. \ref{Fig5}. The isoentropic $C_{ij}$ of ZrC and HfC are greater than those of the isothermal ones for $C_{11}$ and $C_{12}$, while the isothermal $C_{44}$ is equal to the isoentropic one. At 0 K, the differences between the values of $C_{11}^{S}$, $C_{12}^{S}$, $C_{44}^{S}$ from QHA and QSA are quite small ($\sim$0.29 GPa for $C_{11}^{S}$, $~$1.9 GPa for $C_{12}^{S}$, and $~$0.28 GPa for $C_{44}^{S}$). This small difference can (mainly) be attributed to the zero-point vibrational energy considered in the QHA method. At 0 K, $C_{12}^{S}$ and $C_{44}^{S}$ from QHA are 102.9 and 152.5 GPa, respectively, which reasonably agree with the experimental data\cite{chang1966low} (98.4 GPa and 161.1 GPa) and calculation results\cite{liu2014first} (103.5 GPa and 137.8 GPa\cite{liu2014first}). For C$_{11}^{S}$, the value is in good agrement the pervious calculation result\cite{liu2014first} (445.6 GPa) while slightly smaller than the corresponding experimental one\cite{chang1966low} (480 GPa) at 0 K.

For HfC (at 0 K), the calculated values of $C_{11}^{S}$, $C_{12}^{S}$, and $C_{44}^{S}$ from QHA are 500.9, 104.1, and 175.5 GPa, respectively. As temperature increases to 298 K, $C_{11}^{S}$, $C_{12}^{S}$, and $C_{44}^{S}$ become 485.7, 104.1, and 174.4 GPa, respectively. These are in good agreement with the experimental values\cite{weber1973lattice}[$C_{11}$=500 GPa, $C_{44}$=180 GPa at 298 K ($C_{12}$ is not experimentally available)]. Note that the above result are very similar due to the geochemical twins of Zr and Hf. With temperature increasing, $C_{11}^{S}$ of both ZrC and HfC decreases, and the degree of decreasing tendency of $C_{11}^{S}$ calculated from QHA is more prominent than the one calculated from QSA.

\begin{center}
\begin{figure*}
\includegraphics[angle=0,width=0.85\linewidth]{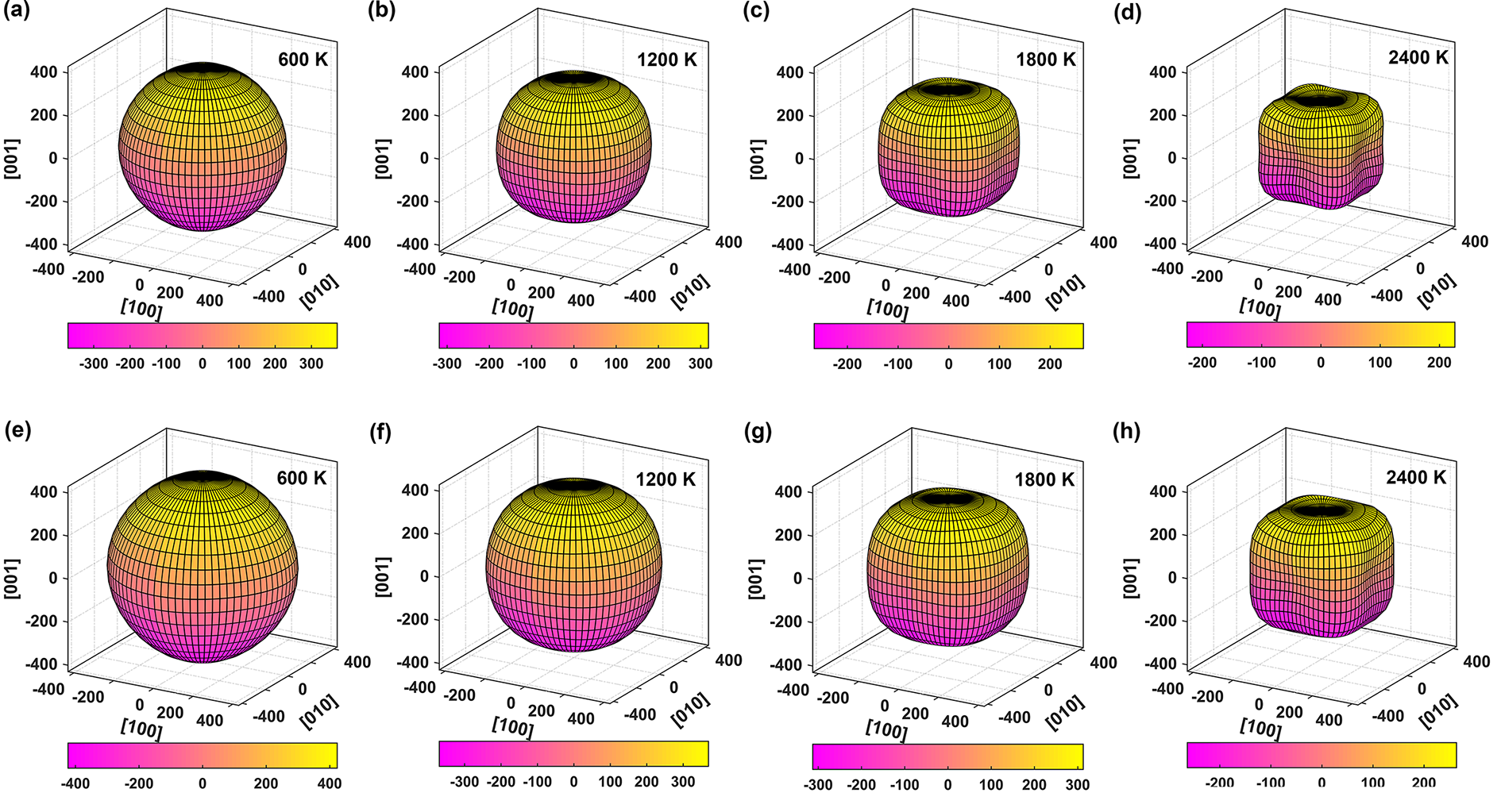}
\caption{\label{Fig7} (Color online) Direction dependence of \emph{E} for (a)--(d) ZrC and (e)--(f) HfC at several temperatures.}
\end{figure*}
\end{center}

\begin{center}
\begin{figure*}
\includegraphics[angle=0,width=0.85\linewidth]{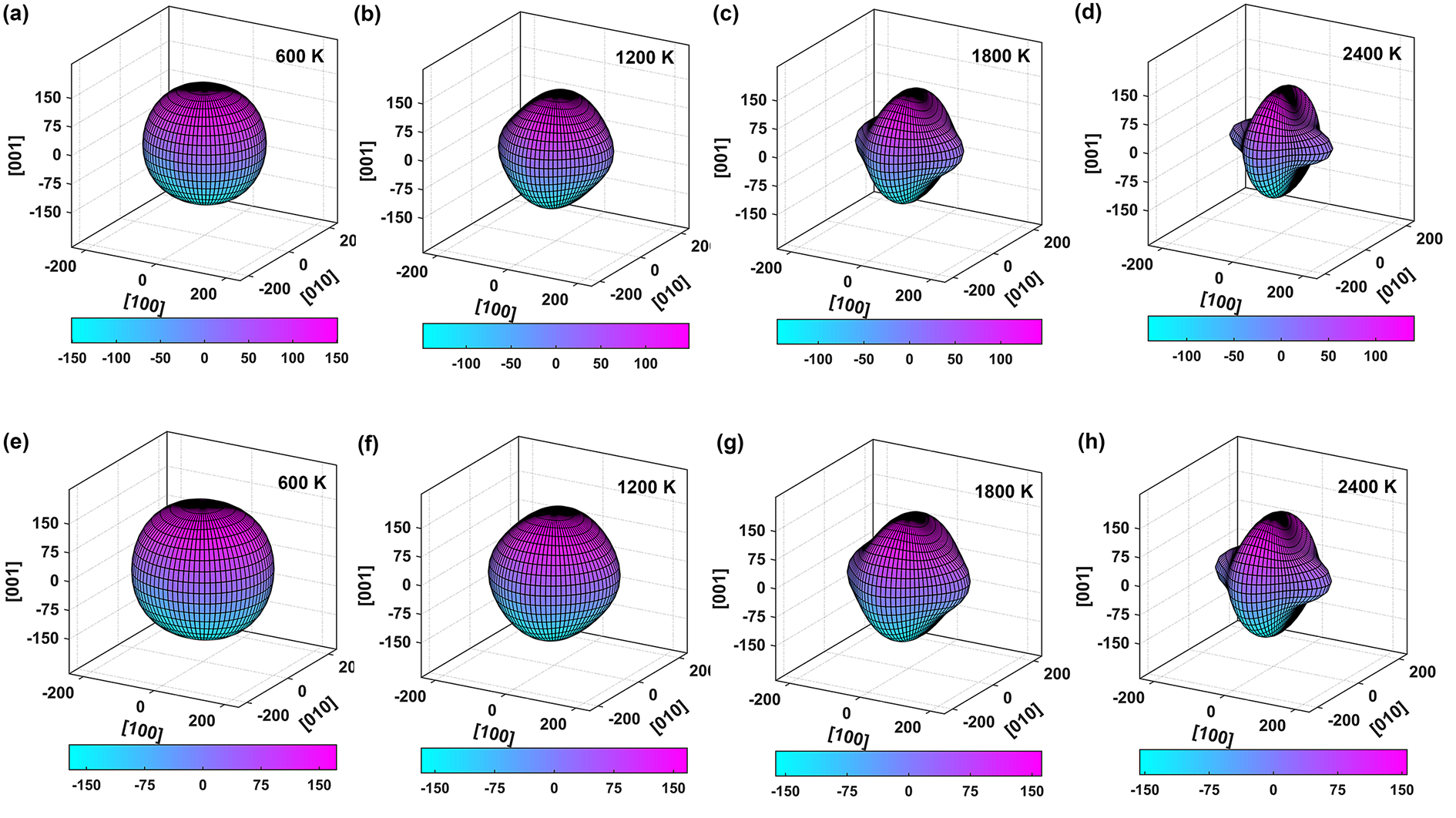}
\caption{\label{Fig8} (Color online) Direction dependence of torsion shear modulus \emph{G}$_{t}$ for (a)--(d) ZrC and (e)--(f) HfC at several temperatures.}
\end{figure*}
\end{center}

As compared above with the experimental data\cite{chang1966low}, the diagonal elastic constant $C_{11}^{S}$ of ZrC under temperature is underestimated either calculating from QHA or QSA, while $C_{12}^{S}$ and $C_{44}^{S}$ from both QHA and QSA are in good agreement. This situation is very similar with a calculation\cite{wang2010first} that the predicted temperature-dependent $C_{11}^{S}$ (from QSA) of cubic Al and Cu are underestimated, while the agreements are good for $C_{12}^{S}$ and $C_{44}^{S}$ (compared with the experimental data\cite{gerlich1969high,kamm1964low,sutton1953variation,chang1966temperature,overton1955temperature}). The same is also found for $\alpha$-Al$_{2}$O$_{3}$\cite{shang2010temperature}, where this underestimation is attributed to stress fluctuations during the QSA calculation. However, the influence of this can be neglected in the calculations herein, since the strain-Helmholtz free energy (instead of the strain--stress method) is used in both QHA and QSA approaches.

According to the isoentropic TDEC calculated from QHA, $C_{11}^{S}$ of ZrC and HfC decrease by 46\% (from 434.8 to 234.9 GPa) and 41\% (from 485.7 to 284.3 GPa), respectively, from room temperature to 2500 K. The variations of $C_{12}^{S}$ and $C_{44}^{S}$ are found to be much smaller. This can be understood as follows, $C_{11}^{S}$ represents the stiffness against a longitudinal strain, which generates a change in volume without a change in shape. The volume change is highly related to the temperature, and thus causes a large change in $C_{11}^{S}$. On the other hand, $C_{12}^{S}$ and $C_{44}^{S}$ are related to the deformation resistance to a transverse expansion (or shear strain), which causes a change in shape without a change in volume. Thus, $C_{12}^{S}$ and $C_{44}^{S}$ are less sensitive of temperature as compared with $C_{11}^{S}$. Note that the quasiharmonic approximation (step 2 in the methods section), which only takes into account the volume dependence of phonon frequencies of lattice vibrations, is only reasonable in a temperature range below the melting point (at temperatures very close to it, phonon--phonon anharmonicity becomes important \cite{grabowski2009ab}).

\begin{center}
\begin{figure*}
\includegraphics[angle=0,width=0.75\linewidth]{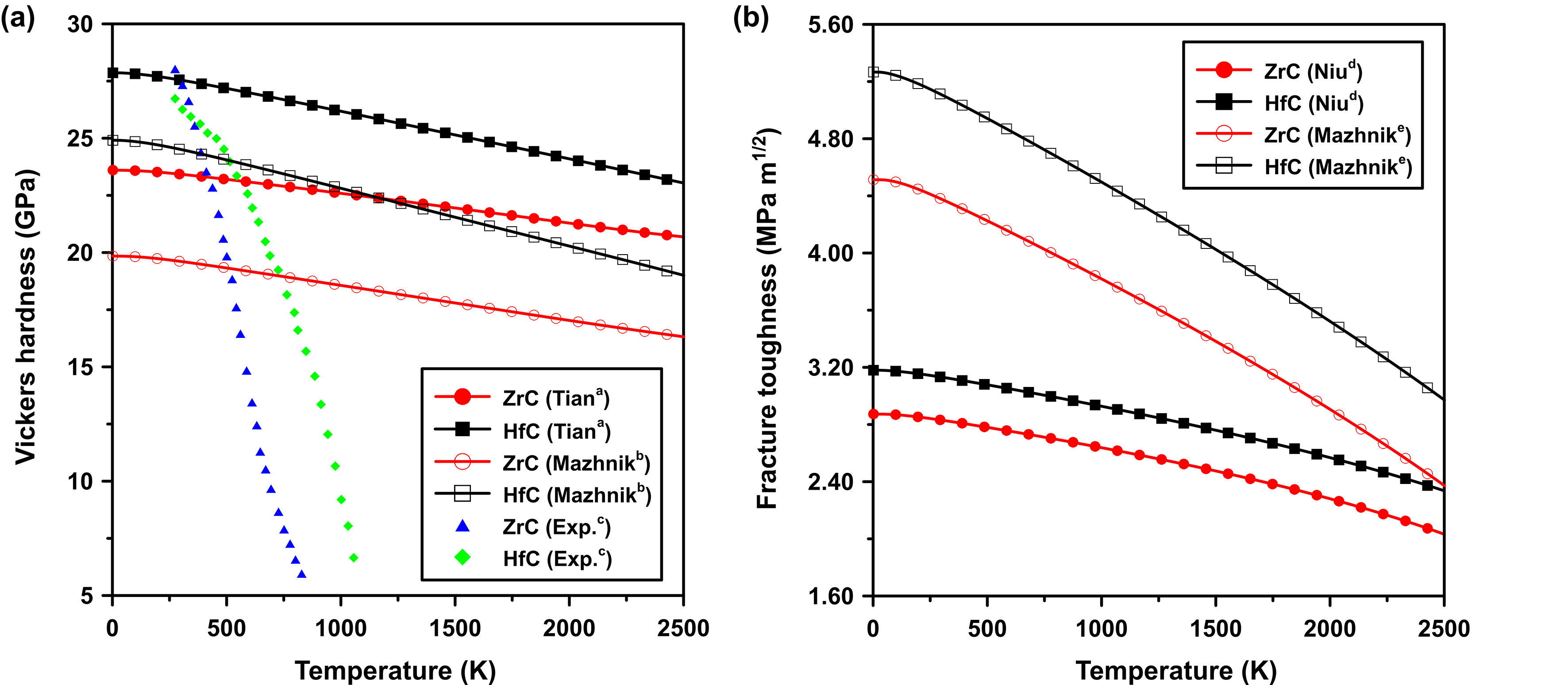}
\caption{\label{Fig9} (Color online) Temperature-dependent (a) hardness and (b) fracture toughness of ZrC and HfC. Tian$^{a}$ is calculated according to the model of Ref. \onlinecite{tian2012microscopic}; Mazhnik$^{b}$, Ref. \onlinecite{mazhnik2019model}; Niu$^{d}$, Ref. \onlinecite{niu2019simple}; Mazhnik$^{e}$, Ref. \onlinecite{mazhnik2019model}, and Exp.$^{c}$ is from Ref. \onlinecite{pierson1996carbides}.}
\end{figure*}
\end{center}

The temperature dependence of elastic moduli is extremely important for the strength of a material under high temperature and thermal shock properties. Based on the calculated isoentropic $C_{ij}$ (as a function of temperature), the isoentropic elastic properties of polycrystalline ZrC and HfC, including bulk modulus \emph{B}, shear modulus \emph{G}, Young's modulus \emph{E}, and Poisson's ratio $\upsilon$, can be evaluated under temperature using Voigt--Reuss--Hill approximation\cite{hill1952elastic} (see Fig. \ref{Fig6}). As expected, the overall tendency is that (temperature-dependent isoentropic) \emph{B}, \emph{G}, and \emph{E} decrease with increasing temperature. The temperature dependence of isoentropic \emph{E} of ZrC obtained using the QHA approach is closer to the experimental result\cite{baranov1973temperature}, as shown in Fig. \ref{Fig6}(c). A comparison of \emph{B}, \emph{G}, and \emph{E} of both ZrC and HfC calculated using QHA and QSA reveals that the difference between the QHA and QSA is small below 250 K, while the decrease of \emph{B}, \emph{G}, and \emph{E} estimated from QHA is more significant than these from QSA above 250 K. For example, as the temperature increases from 298 to 2500 K, \emph{B}, \emph{G}, and \emph{E} of ZrC decrease by 38\%, 30\%, 32\% according to the QHA method, whereas they only decrease by 25\%, 16\%, 18\% from QSA. The value of \emph{G}/\emph{B} can reflect the brittleness (ductility) of a material; a material is deemed to be brittle if \emph{G}/\emph{B} $>$ 0.57 and less than this indicates a ductile material\cite{pugh1954xcii}. At 298 K, the calculated isoentropic \emph{G}/\emph{B} ratio (from QHA) of ZrC and HfC is 0.74 and 0.78, respectively, indicating a behavior of brittleness and their brittle characteristic are enhanced at high temperatures based on the calculation.

All of the crystals are elastically anisotropic, which can be favorable for the emergence of microcracks\cite{ravindran1998density}. Calculating and visualizing the elastic anisotropy is important for understanding such properties, and optimizing them for practical applications. Through analyzing the calculated temperature-dependent isoentropic $C_{ij}$ (from QHA), the directional dependence of \emph{E} and torsion shear modulus \emph{G}$_{t}$ for ZrC and HfC at several temperatures are displayed in Figs. \ref{Fig7} and \ref{Fig8}, respectively. The degree of elastic anisotropy in a system can be reflected by observing the amount of deviation of the shape of these quantities from a sphere. As temperature increases, the extent of anisotropy of \emph{E} and \emph{G}$_{t}$ for both ZrC and HfC increases. Within the same temperature range, the magnitudes of anisotropy of \emph{E} and \emph{G}$_{t}$ for ZrC are larger than that of HfC.

Hardness of a materials is defined as its ability to resist plastic deformation, and fracture toughness \emph{K}$_{IC}$ describes the resistance of a material against crack propagation\cite{ashby1993materials}. The trend of both of these are important from an application standpoint. Based on the temperature-dependent isoentropic $C_{ij}$ (from QHA), temperature-dependent vickers hardnesses can be predicted according to the Tian's\cite{tian2012microscopic} and Mazhnik's\cite{mazhnik2019model} models. Fracture toughness is estimated using Niu's\cite{niu2019simple} and Mazhnik's\cite{mazhnik2019model} ones. For both ZrC and HfC at 298 K, the hardnesses calculated from Chen's model agree well with experimental results\cite{pierson1996carbides}, while the ones calculated from Mazhnik's model are a little bit lower. However, for both ZrC and HfC, with increase of temperature, the estimated hardnesses from both models are vastly overestimated, as shown in Fig. \ref{Fig9}(a). One possible explanation is that the experimentally-tested hardness of ZrC and HfC under temperature may vary considerably depending on the composition and the presence of impurities of the samples\cite{pierson1996carbides}. Another is that these empirical hardness models might fail to take account the dislocation propagation in a material, which significantly have a negative influence on the hardness of a material as temperature increases.

At 298 K, the calculated fracture toughnesses of ZrC and HfC from Niu's model are 2.82 and 3.12 MPa m$^{1/2}$, respectively,  and from Mazhnik's model are 4.38 and 5.11 MPa m$^{1/2}$. Both decrease with increasing temperature, and the descending rate estimated from Mazhnik's model is higher than that of from Niu's one (e.g. \emph{K}$_{IC}$ of ZrC decreases 33.6\% from Mazhnik's model and 23.4\% from Niu's one from 298 to 2000 K). Note that one should be careful about the capacity of using these models to predict the temperature-dependence though, since it might ignore behavior of crack propagation under temperature. Indeed, the mechanism controlling the changes in the hardness and fracture toughness of a material at high temperatures is actually more complex, and the current models to calculate them should to be re-evaluated.

\section{Conclusions}

Isothermal and isoentropic elastic constants for cubic ZrC and HfC have been calculated using qusi-harmonic and qusi-static approaches over a wide range of temperature. Significant differences were found between the two approaches. This suggests a surprising importance of the vibrational component of the free energy to calculate mechanical properties, even for systems such as these with heavy elements. With increasing temperature, the isoentropic $C_{11}$ and $C_{44}$ gradually decrease, while $C_{12}$ (slightly) increases for both ZrC and HfC. For both, there is a also decrease of bulk modulus \emph{B}, shear modulus \emph{G}, and Young's modulus \emph{E} at high temperatures. The temperature-dependent elastic properties of HfC is comparatively superior to those of ZrC. Besides, Young's and the torsion shear moduli of both ZrC and HfC becomes substantially more anisotropic at high temperatures. The validity of using hardness and fracture toughness models to estimate their temperature-dependence was discussed. These results are expected to have important applications in applied physics, and also for future theoretical and computational work.

\section{acknowledgments}
J.\ M.\ M.\ acknowledges startup support from Washington State University and the Department of Physics and Astronomy thereat. J.\ Z.\ thanks Xiao Dong (Nankai University) and Xinfeng Li (Sun Yat-sen University) for valuable discussion.

\end{document}